# Uncovering the nutritional landscape of food

Seunghyeon Kim[a,b], Jaeyun Sung[a], Mathias Foo[a], Yong-Su Jin[c,d], Pan-Jun Kim[a,b,1]

[a]Asia Pacific Center for Theoretical Physics, Pohang 790-784, Republic of Korea

[b]Department of Physics, Pohang University of Science and Technology, Pohang 790-784, Republic of Korea

[c]Department of Food Science and Human Nutrition, University of Illinois at Urbana-Champaign, Urbana, IL 61801

[d]Institute for Genomic Biology, University of Illinois at Urbana-Champaign, Urbana, IL 61801

[1]Corresponding author. E-mail: pjkim@apctp.org

**Recent progresses in data-driven analysis methods, including network-based approaches, are revolutionizing many classical disciplines. These techniques can also be applied to food and nutrition, which must be studied to design healthy diets. Using nutritional information from over 1,000 raw foods, we systematically evaluated the nutrient composition of each food in regards to satisfying daily nutritional requirements. The nutrient balance of a food was quantified and termed nutritional fitness; this measure was based on the food's frequency of occurrence in nutritionally adequate food combinations. Nutritional fitness offers a way to prioritize recommendable foods within a global network of foods, in which foods are connected based on the similarities of their nutrient compositions. We identified a number of key nutrients, such as choline and α-linolenic acid, whose levels in foods can critically affect the nutritional fitness of the foods. Analogously, pairs of nutrients can have the same effect. In fact, two nutrients can synergistically affect the nutritional fitness, although the individual nutrients alone may not have an impact. This result, involving the tendency among nutrients to exhibit correlations in their abundances across foods, implies a hidden layer of complexity when exploring for foods whose balance of nutrients within pairs holistically helps meet nutritional requirements. Interestingly, foods with high nutritional fitness successfully maintain this nutrient balance. This effect expands our scope to a diverse repertoire of nutrient-nutrient correlations, which are integrated under a common network framework that yields unexpected yet coherent associations between nutrients. Our nutrient-profiling approach combined with a network-based analysis provides a more unbiased, global view of the relationships between foods and nutrients, and can be extended towards nutritional policies, food marketing, and personalized nutrition.**

# Introduction

Among the many factors that influence our choice of food for consumption, such as palatability, financial costs, and cultural background [1–4], nutritional sufficiency is given the highest priority for the maintenance of human health [5]. Therefore, in response to public concerns regarding wellness, considerable efforts have been made to accumulate nutritional knowledge, e.g., the nutritional composition of foods, the health consequences regarding the intake of particular nutrients, and the recommended levels of nutrient consumption [6, 7]. The nutritional data accumulated from these efforts have been applied to various practical purposes, such as the design of dietary recommendations [8], the formulation of optimal livestock feed [9, 10], and the ranking of foods based on their nutrient content [11, 12]. These studies have certainly served a significant role in addressing many practical concerns regarding nutrition and diet. Nevertheless, prominent systematic and comprehensive analyses of foods and their nutrients remain lacking, which provides a clear opportunity to elicit new scientific insight and thereby broaden the impact of previously accumulated nutritional data.

Data-driven analysis methods, including network-based approaches, are now widely used for fundamental quantitative inquiries regarding various complex biological, technological, and social systems [13–16]. Such techniques have also been applied to foods: a recent analysis of a network that connects various food ingredients to flavor compounds revealed unforeseen regional variations in culinary cultures [17]. Despite this study on the global connections between food ingredients and flavor compounds, to the best of our knowledge, there has not been any work that utilizes comprehensive network-related approaches on only raw foods and their nutrients. Herein, we present an unprecedented global view of the relationships between foods and nutrients through a systematic analysis of a publicly available food and nutritional dataset. We develop a unique quantification system to measure the nutritional adequacies of various foods and identify the key elements, which can then be interpreted in the context of network patterns among foods and nutrients. The results from this analysis not only help improve our basic understanding of the nutritional structure of the human diet, but also have a wide range of implications for nutritional policies, the food industry, and personalized nutrition.

## Results and Discussion

### Hierarchical Organization of the Food-Food Network

We started by constructing a food-food network composed of various raw foods connected by weighted links. In this study, "raw foods" indicate raw foods, as well as other foods with minimally-modified nutrient contents, e.g., frozen and dried foods (Supplementary Information). The number of raw foods was initially 1,068 and we systematically unified foods redundant in their nutrient contents, giving rise to a total of 654 foods in the network (see Supplementary Information). The weight of each link that connects the two foods represents the similarity of the foods' nutritional compositions (Supplementary Information). For example, in this network, persimmon and strawberry have very similar nutritional compositions, especially in their relative amounts of calcium, potassium, vitamin C, phosphorus, amino acids, and fat ($P = 1.1 \times 10^{-12}$). Fig. 1(a)–(c) shows a global architecture of the food-food network, clearly revealing its multi-scale organization wherein nutritionally similar foods are recursively grouped into a hierarchical structure. At the highest level of the organization, the network can be largely divided into two parts, the animal-derived part and the plant-derived part (Fig. 1(a)). The animal-derived part consists of foods that mostly have large amounts of proteins and/or fats relative to the amounts of carbohydrates, such as fish, meat, and eggs. In contrast, the plant-derived part contains foods that generally have small amounts of proteins, such as fruits, grains, mushrooms, and vegetables (with the exception of a few foods such as alfalfa seeds, in which protein constitutes 55.6% of the dry weight). Within the animal-derived part, we identified several foods whose nutrients were similar to those within the plant-derived part (and vice versa); these foods thus serve as interesting bridges across the two large clusters. One example of pairs of 'bridge' foods is northern pike liver and sprouted radish seeds, which have similar nutritional compositions, especially in their relative amounts of fat, iron, and niacin ($P = 0.009$; see Supplementary Information).

At a deeper level of the hierarchical structure of the food-food network (i.e., within either the animal- or plant-derived cluster), we found that foods can be grouped amongst each other, again according to their relative levels of macronutrients, i.e., proteins, fats, and carbohydrates (Supplementary Information). From this observation, we identified two categories within the animal-derived foods: the protein-rich category (209 foods, such as fish, meat, and poultry) and the fat-rich category (7 different animal fats from pork, lamb, beef, and veal). In addition, the plant-derived foods were primarily divided into three large categories: the fat-rich category (34 nuts, seeds, avocados, and rice bran), the carbohydrate-rich category (208 fruits, grains, root vegetables, seaweeds, and others), and the low-calorie category (186 vegetables, spices, herbs, mushrooms, and others). A fat-rich category is found in both animal-derived and plant-derived foods; however, the foods that belong to one of the fat-rich categories are largely distinguishable from the foods of the other category by their abundance of saturated fatty acids (animal fats generally contain much more saturated fatty acids than plant fats. See Supplementary Information).

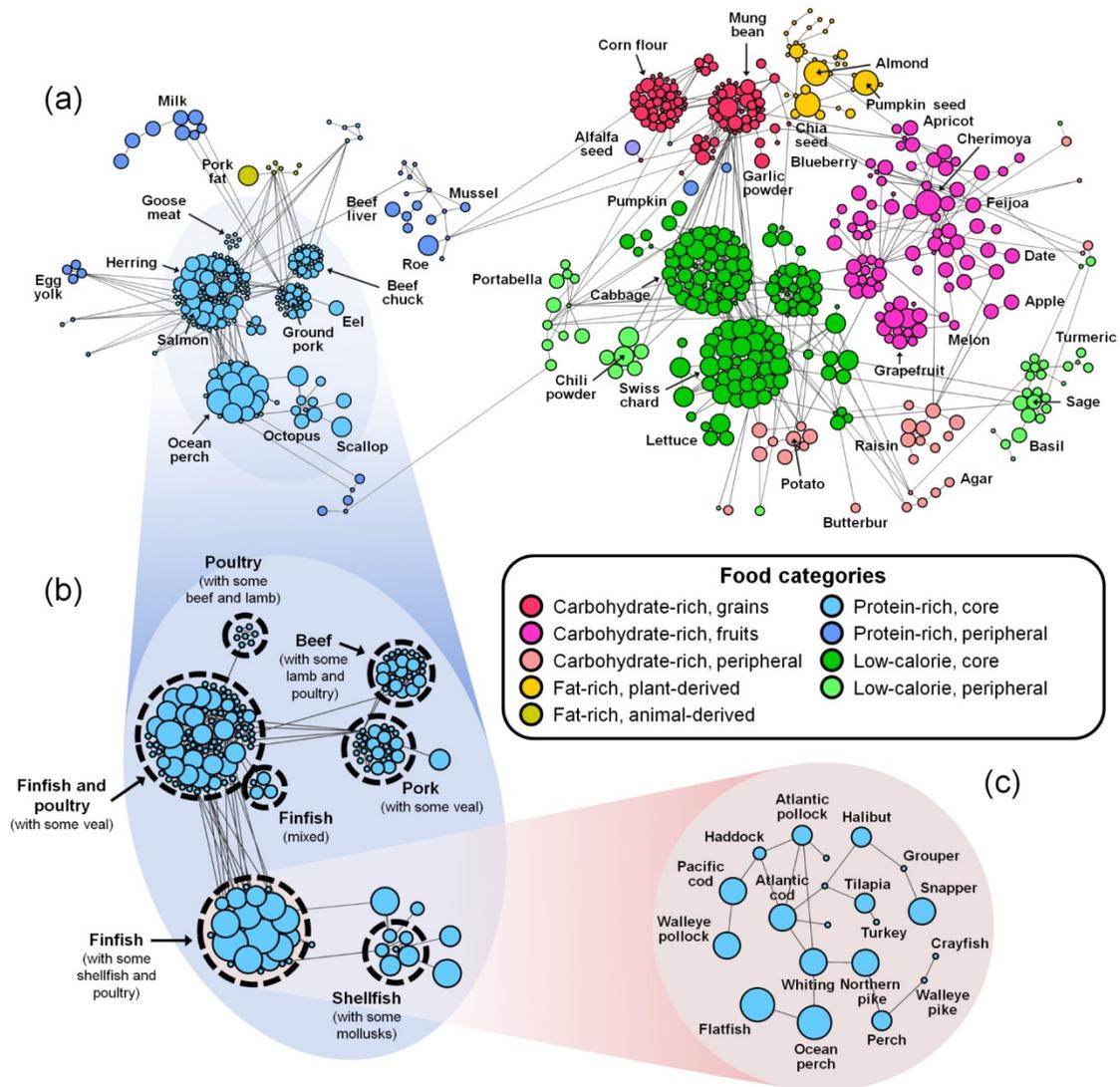

**Fig. 1. The food-food network. (a–c)** Large-scale to small-scale overviews of the network. Each node represents a food, and nodes are connected through links that reflect the similarities between the nutrient contents of foods. The network in (a) is composed of animal-derived (left) and plant-derived (right) foods. A part of the animal-derived foods is magnified in (b), which shows seven different clusters of foods. The members of one of these clusters, the cluster 'Finfish (with some shellfish and poultry)', are shown in (c). In (a)–(c), each node is colored according to the food category. The size of each node corresponds to the nutritional fitness (NF) of the food (Fig. 2(a) and 2(b)). For visual clarity, we only show the topologically-informative connections between the foods (represented by links with the same thickness), and we omit six foods that have loose connections to the network (see Supplementary Information for details).

Finally, to attain the finest level of the organization, we continued our hierarchical clustering approach (grouping foods with similar nutrient contents) for all foods. We discovered that the global network structure is predominantly composed of 41 distinct food clusters, which encompass 76.9% of the total foods (Supplementary Information). Among those food clusters, more than half of the clusters (22 clusters) include less than six foods each, but there are also a significant number of clusters (11 clusters) that include more than ten foods each. Fig. 1(b) shows several clusters that primarily include finfish, shellfish, beef, pork, and poultry. In general, the organismal sources of the foods in each cluster were homogenous or similar based on their phylogenetic lineage. However, we faced a few cases without this trend. Finfish and poultry belonged to the same two clusters, as illustrated in Fig. 1(c) wherein turkey does exist in the finfish-majority cluster. This unexpected result is accounted for by the fact that turkey and tilapia share similar relative proportions of various amino acids, minerals, cholesterol, and niacin ($P = 1.3 \times 10^{-10}$). Overall, from the coarse to fine scales, the global structure of our food-food network not only exhibits hierarchical patterns consistent with common nutritional knowledge, but also discloses unexpected relationships between foods clearly portrayed by our unbiased methodology.

**Characterization of Nutritional Fitness**

The food-food network provides a global view of the nutritional connections between foods; however, we desire more direct information on which foods can lead to good health outcomes. Specifying food quality based on nutrient contents will help consumers meet the nutrient intakes necessary for good health.

Suppose a hypothetical scenario wherein an ideal food contains all necessary nutrients to meet, but not exceed, our daily nutrient demands. In this case, consuming only this food, without any other food, will provide the optimal nutritional balance for our body. In the absence of such a prime, ideal food, a realistic alternative would be to consume a set of foods, small in number, that still satisfies nutritional recommendations (in fact, we find that the minimum set consists of four different raw foods. See Supplementary Information). Using this concept, we examined all potential food combinations and identified sets that have the smallest numbers of different foods and meet our daily nutrient demands in each entirety. We henceforth call these food sets the *irreducible food sets*. Foods with a frequent occurrence across these combinations are likely to provide very balanced nutrients (the number of different foods that comprise an irreducible food set was limited to a small value. Otherwise, it is hard to estimate the true nutritional adequacy of the foods, solely examining the frequency of the food occurrence across the food sets. For example, a nutritionally-poor food in a certain set, without any proper limitation on the set size, might be easily complemented by many other foods in the same set, to meet our daily nutrient demands). To characterize the nutritional adequacy of foods, we introduce the *nutritional fitness* ($NF_i$) measure; this value monotonically increases with the number of irreducible food sets that include food $i$, and the value of $NF_i$ ranges from zero to one (Fig. 2(a)). A large $NF_i$ suggests that food $i$ is nutritionally favorable. In this work, we considered the nutritional requirement of a physically active 20-year-old male and constrained the total weight of daily food consumed, when the $NF_i$ of each food was calculated (Supplementary Information). Although profiling and scoring foods based on the nutrient contents have been attempted in many previous studies [11, 12], in general, their methods involve rather arbitrarily-structured mathematical formulas and explicit weighting factors, which may lead to possibly biased results. In contrast, our study takes a conceptually different, clearly defined approach to prioritize foods that are nutritionally adequate based on the outputs of optimization problems in which all nutrient levels are simultaneously constrained within the ranges recommended for daily intake.

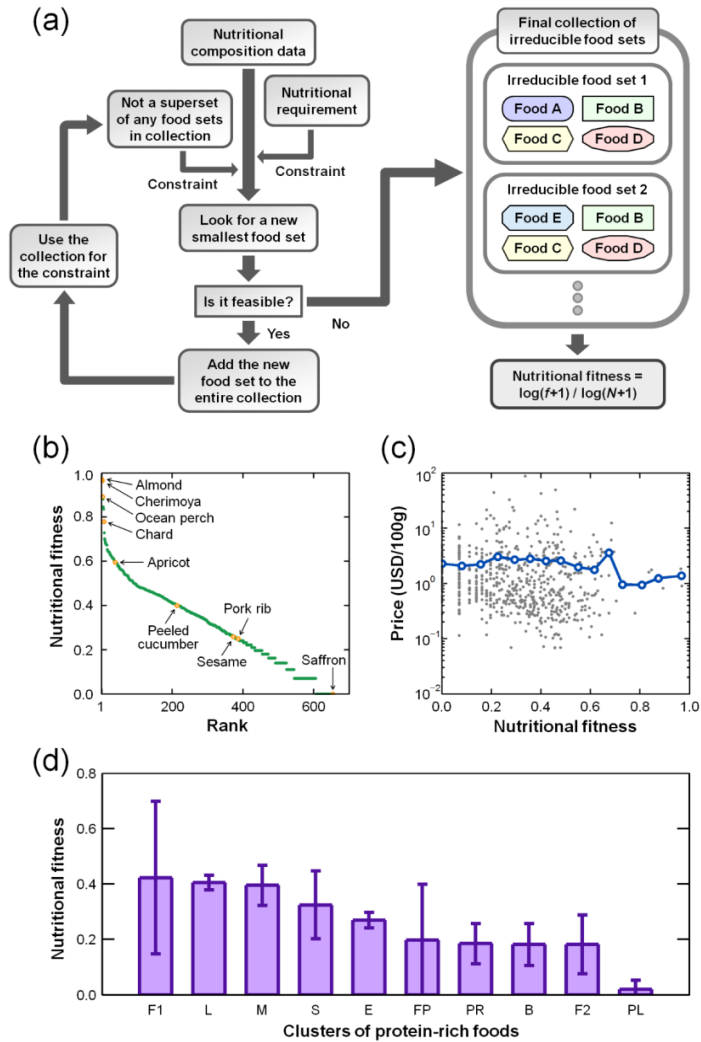

**Fig. 2. Characteristics of nutritional fitness (NF). (a)** Flow chart for calculating NF. See Supplementary Information for the detailed procedures of the flow chart. At the end, we assign NF = log($f$+1)/log($N$+1) to each food, where $f$ is the number of irreducible food sets that include the food, and $N$ is the number of all irreducible food sets. An irreducible food set is defined as a set of different foods that satisfies the following two conditions: it meets our daily nutrient demands in its entirety, and no set is a superset of another set. We limit the number of different foods in each irreducible food set and the total weight of foods therein (Supplementary Information). A large NF suggests that the food is nutritionally favorable. **(b)** NFs of foods, sorted in descending order. **(c)** NF versus price (per weight) for each food (gray). The blue line indicates the average prices along NFs. **(d)** NFs of foods (average and standard deviation) in each food cluster of the protein-rich category. Clusters are abbreviated as follows. F1: Finfish (with some shellfish and poultry); L: Animal liver; M: Milk; S: Shellfish (with some mollusks); E: Eggs; FP: Finfish and poultry (with some veal); PR: Pork (with some veal); B: Beef (with some lamb and poultry); F2: Finfish (mixed); PL: Poultry (with some beef and lamb).

From our calculations, then, which foods have the highest NFs? The three foods with the highest NFs were almond, cherimoya, and ocean perch, which had NF values of 0.97, 0.96, and 0.89, respectively (Fig. 2(b); NF = 0.30 ± 0.19 for all foods). Almond, which is the food with the highest NF, belongs to a fat-rich category in the food-food network, whereas cherimoya and ocean perch belong to the carbohydrate-rich and protein-rich categories, respectively. An interesting question is whether foods with high NFs tend to be more expensive to purchase than foods with low NFs. Fig. 2(c) shows essentially no correlation between a food's NF and price per weight ($r = -0.02$, $P = 0.65$; see also Supplementary Information).

One important issue here is whether the categories to which the foods belong (delineated in our food-food network) play any role in the NF-driven prioritization of foods. An equivalent viewpoint of this issue is to ask whether the comparison of NFs should be made across all foods included in our study or rather in a category-specific manner. Regarding this issue, we found that most irreducible food sets are composed of foods that cover all four major categories (i.e., protein-rich, fat-rich, carbohydrate-rich, and low-calorie categories) most likely because foods from different categories independently contribute to satisfying the overall nutritional requirements of our diet. In this sense, a food found in an irreducible food set and belonging to a particular category cannot be easily replaced by a food from another category without compromising the food set's entire nutritional adequacy. However, a different food from the same category is allowed to serve as a replacement. Therefore, using NFs to prioritize foods for nutritionally-balanced diets should only be done for foods that belong to the same category.

As previously discussed, the four major categories in the food-food network were further divided into many finer-scale food clusters. Between foods from different clusters of the same category, we found that their NFs provide a moderately distinguishing characteristic: in the protein-rich category, foods that belong to the finfish, animal liver, and milk clusters had higher NFs, on average, than foods in the pork, beef, and poultry clusters (Fig. 2(d); one exception is a finfish cluster with a relatively low NF, but this cluster only contains approximately 6% of all finfish). In the fat-rich category, nuts and seeds tend to have higher NFs than animal fats (Supplementary Information). In the carbohydrate-rich category, fruits tend to have higher NFs than grains and legumes (Supplementary Information). In the low-calorie category, vegetables and peppers have higher NFs than herbs and spices (Supplementary Information). Thus, our systematic analysis using NFs offers a prioritized list of foods from each of the major food categories.

**Bottleneck Nutrients: Key Contributors to High Nutritional Fitness**

The NFs of foods in our study were found to be widely dispersed. An interesting avenue to pursue moving forward would be to more deeply examine the identities of the individual nutrients; specifically, what particular nutrients significantly influence the NF of the food?

For example, in the case of the almond, what nutrients were responsible for this food having the highest NF in the fat-rich category? In order to identify these key nutrients, we initially substituted high-NF foods from irreducible food sets with low-to-moderate-NF foods from the same major category. Next, we inspected which nutrient levels in the whole irreducible food set were significantly altered to dissatisfy daily requirements. We interpreted these sets of nutrients as the main contributors to the foods' high NF values; thus we refer to these nutrients as the *bottleneck nutrients* for high NF (Supplementary Information).

**Table 1. Examples of bottleneck nutrients for high nutritional fitness (NF)**

| Food category | Nutrient name | Remark |
| --- | --- | --- |
| Protein-rich | Choline | Favorable for NF |
|  | Vitamin D | Favorable for NF |
|  | Total lipid | Unfavorable for NF |
|  | Cholesterol | Unfavorable for NF |
| Fat-rich | Linoleic acid | Favorable for NF |
|  | Choline | Favorable for NF |
|  | Manganese | Unfavorable for NF |
| Carbohydrate-rich | Carbohydrate | Favorable for NF |
|  | α-Linolenic acid | Favorable for NF |
|  | Manganese | Unfavorable for NF |
|  | Folate | Unfavorable for NF |
| Low-calorie | Choline | Favorable for NF |
|  | α-Linolenic acid | Favorable for NF |

For each food category, we list the two most favorable and two most unfavorable bottleneck nutrients based on the regression coefficients (Supplementary Information). If the total number of favorable or unfavorable bottleneck nutrients for a given food category was less than two, we listed all. The full list of bottleneck nutrients is available in Supplementary Information, which indicates choline is a favorable bottleneck nutrient in every food category.

Table 1 presents examples of bottleneck nutrients, which can be classified into two types. The first type is nutrients that are not sufficiently found in many low-to-moderate-NF foods. The presence of these nutrients can thus be considered a *favorable* condition for foods to have high NF values. Linoleic acid is one of such favorable nutrients for foods from the fat-rich category. The daily recommendation for this fatty acid is approximately 5-10% of the total calorie intake. However, surprisingly, 90.2% of all fat-rich foods do not contain this important nutrient. A notable exception is almond (the food with the highest NF in the fat-rich category), which had as much as 12.1 g/100 g of linoleic acid. The second type of bottleneck nutrients is found much more abundantly in many low-to-moderate-NF foods; thus, this type is *unfavorable* for increasing a food's NF. In the protein-rich category, cholesterol is one of such unfavorable bottleneck nutrients. We found that dried nonfat milk, which is ranked in the top 12% of foods with the highest NFs in this category, has 20 mg/100 g of cholesterol. This amount is 5.1 times less than the average cholesterol content (102 mg/100 g) in other protein-rich foods. In the carbohydrate-rich food category, α-linolenic acid and manganese are favorable and unfavorable bottleneck nutrients, respectively. Cherimoya, the food with the highest NF in this category, has 28.3 times more α-linolenic acid (159 mg/100 g) and 10.6 times less manganese (93 μg/100 g) than all other carbohydrate-rich foods on average. Furthermore, in this category, folate was identified as an unfavorable bottleneck nutrient, despite being a well-known essential vitamin. This occurs because most carbohydrate-rich foods (91.8% of all foods in this category) contain a rather large amount of this nutrient (101.6 ± 157.4 μg DFE/100 g); therefore, consumption of these foods can cause the total folate intake to easily exceed the daily recommended levels when consumed with foods of other categories. Some foods in our analysis may have been fortified with folate; however, we could not identify the clear evidence solely from our dataset (Supplementary Information).

An interesting question to raise here is why certain types of foods in the same category have noticeably different NF values. For example, in the protein-rich category, finfish tend to have a higher NF than poultry (Fig. 2(d)), despite similarities in their overall nutrient compositions ($P < 2.0 \times 10^{-5}$). We found that choline, a favorable bottleneck nutrient essential for normal body functioning [18], was substantially more abundant in finfish (Supplementary Information). Other bottleneck nutrients that happen to separate foods, especially those from different clusters within the same food category, are shown in Supplementary Information. Our results therefore imply that specific bottleneck nutrients can play a critical role in the discrepancy between the high- and low-NF foods of a given food category.

Among all bottleneck nutrients from each of the four major food categories, we found choline to be a favorable bottleneck nutrient in every category. This nutrient is an important factor for a wide range of physiological processes, from cell membrane synthesis to neurotransmitter metabolism, and its deficiency is now thought to have an impact on a number of diseases [18, 19]. Among all foods in our study, 61.2% of them provide choline to varying degrees. However, the choline contents of these foods are generally insufficient to satisfy the daily recommended level (minimum intake of 550 mg); for half of these foods, the choline content is less than 30.9 mg/100 g. For this reason, we believe that choline was noticeable in a collection of foods with high NFs across all major food categories. Considering a degree of the uncertainty in the dietary requirement for choline, which may be related to genetic polymorphisms [18], it will be valuable to further examine the effects of the altered requirement for choline in our analysis. Finally, we suggest that deeper analyses into these distinguishing bottleneck nutrients may be warranted when the prioritization of foods is of interest.

**Synergistic Bottleneck Effects of Nutrient Pairs**

The fact that specific nutrients can either enhance or diminish the NF of foods encourages us to examine beyond the effect of a single nutrient and to determine whether multiple nutrients, when considered together, can exert such characteristics. In this regard, consider the strategy for how we discovered bottleneck nutrients. Briefly, within irreducible food sets, a high-NF food was systematically replaced with low-to-moderate-NF foods; the nutrients for which the daily requirement was no longer met, as a direct result of these replacements, were subsequently identified. Analogously, one can investigate this same attribute from *pairs* of nutrients. Specifically, when a high-NF food is replaced, the resulting quantity of either of two nutrients in a pair (the quantity from the whole irreducible food set) may no longer meet their respective recommended intake levels. In our collection of irreducible food sets, we found that not only indeed do these pairs of nutrients exist but also can occur more frequently than expected by chance when each of the two nutrients is considered separately. Thus, this result serves as direct evidence of the synergistic bottleneck effect, which is simultaneously produced by pairs of nutrients and contributes to a high NF of foods.

We now introduce $\Phi_{ij}^k$, which is a measure of the degree of such synergism between two nutrients $i$ and $j$ for the high NF of food $k$ (see Supplementary Information). Supplementary Information presents the list of synergistic nutrient pairs with large $\Phi_{ij}^k$s. In the case of choline and cholesterol, this nutrient pair exhibits strong synergism in ocean perch ($\Phi_{ij}^k = 22.0$, $P < 10^{-16}$), the highest-NF food among all foods in the protein-rich category. Previously, we found that choline and cholesterol are favorable and unfavorable bottleneck nutrients, respectively, in the foods of this category. Our analysis demonstrates that, when favorable and unfavorable nutrients were found in highly synergistic bottleneck pairs, in general, their quantities tended to be positively correlated across the foods in each of the four major categories (Fig. 3; $P < 2.0 \times 10^{-4}$ to $P = 0.04$). This positive correlation, which was identified among nutrients that have contradicting roles in influencing NF, contributes to the previously discussed difficulty in maintaining nutrient balance, i.e. simultaneously meeting the respective daily nutritional requirements, in irreducible food sets.

Intriguingly, the individual nutrients in a pair that exhibits a synergistic bottleneck effect are not necessarily bottleneck nutrients themselves that can separately impact the NFs of foods. For example, vitamin E and folate constitute a synergistic nutrient pair that contributes to the high NF in almond among fat-rich foods ($\Phi_{ij}^k = 10.5$, $P < 10^{-16}$). These two nutrients are not bottleneck nutrients in the fat-rich category; however, vitamin E and folate are moderately favorable and unfavorable for high NF, respectively, and they do share a positive correlation in their abundances across fat-rich foods. Almond, the highest-NF food in the fat-rich category, has 7.6 times more vitamin E (26.2 mg/100 g) and 3.1 times less folate (50 μg DFE/100 g) than expected from the overall trend of the fat-rich foods that have positively-correlated quantities ($r = 0.34$) of the two nutrients. Furthermore, in the case of flatfish (which has the second highest NF among protein-rich foods), vitamin B12 (1.1 μg/100 g) and folate (5.0 μg DFE/100 g) comprise a synergistic bottleneck pair ($\Phi_{ij}^k = 10.3$, $P < 10^{-16}$), although neither of the two nutrients are bottleneck nutrients themselves in the protein-rich foods. Table 2 shows the full list of synergistic pairs that have non-bottleneck nutrients. These

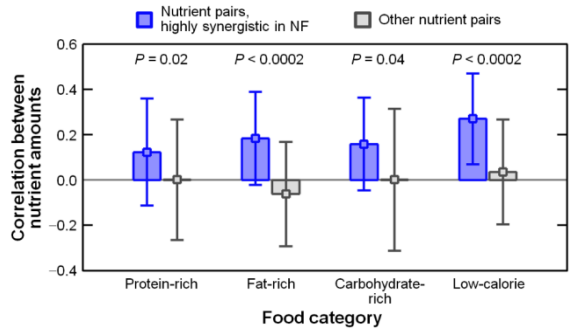

**Fig. 3. Correlation between abundances of two nutrients (one nutrient is favorable and the other nutrient is unfavorable for NF) across foods in each food category.** For highly synergistic nutrient pairs ($\Phi_{ij} > 2.0$; blue) and the other pairs ($\Phi_{ij} \leq 2.0$; grey), we present the respective averages and standard deviations of the correlations (see Supplementary Information).

**Table 2. Synergistic bottleneck pairs for high NF, which are composed of non-bottleneck nutrients**

| Food category | Nutrient 1 | Nutrient 2 | Food | Remark |
|---|---|---|---|---|
| Protein-rich | Vitamin B12 | Folate | Flatfish | F, U |
| | Vitamin B12 | Linoleic acid | Flatfish | F, F |
| Fat-rich | Carbohydrate | Folate | Almond | F, U |
| | Vitamin E | Niacin | Almond | F, U |
| | Vitamin E | Folate | Almond | F, U |
| | Vitamin E | Iron | Almond | F, U |
| | Carbohydrate | Niacin | Almond | F, U |
| | Vitamin E | Sodium | Almond | F, U |
| | Folate | Total lipid | Almond | U, U |
| | Folate | Saturated fat | Almond | U, U |
| | Niacin | Total lipid | Almond | U, U |
| Carbohydrate-rich | Vitamin E | Total lipid | Tangerine | F, U |
| | Calcium | Iron | Kumquat | F, U |

For each food category, we list synergistic bottleneck pairs ($\Phi_{ij} > 2.0$) composed of nutrients (in the second and third columns) that are not bottleneck nutrients themselves for high NF in that food category. Only food in which a given pair of nutrients exhibits the strongest synergism (among multiple foods) for high NF is shown in the fourth column. In the fifth column, 'F' ('U') denotes that the nutrient is 'favorable' ('unfavorable') for a high NF of the food in the fourth column (see Supplementary Information). For example, 'F, U' indicates that a nutrient in the second column is favorable, whereas the nutrient in the third column is unfavorable. This table shows only the cases with a definite 'F' or 'U' (Supplementary Information). Foods in the low-calorie category do not have synergistic pairs of non-bottleneck nutrients.

results evince the fact that balancing multiple nutrients simultaneously cannot be as simple as expected from balancing individual nutrients. Therefore, the study raises the importance of nutrient-to-nutrient connections in the context of balancing multiple nutrients simultaneously, which adds another layer of complexity when understanding the nutritional adequacy of foods.

**The Nutrient-Nutrient Network**

In light of the synergistic bottleneck effects, the previously discussed nutrient-nutrient correlations across foods extend our interest to a comprehensive picture of the associations between nutrients. In this aspect, we performed an extensive, unbiased survey of these nutrient-nutrient correlations by constructing a nutrient-nutrient network, in which nodes are nutrients, and nutrients are connected to each other through correlations in their abundances across foods. For illustration, Fig. 4 presents the nutrient-nutrient network based on the correlations across all foods (we also consider correlations measured in a food-group-specific manner for subsequent analyses). In our network, glucose and fructose are examples of nutrients that are connected through a large correlation ($r = 0.85$, $P = 7.4 \times 10^{-23}$). Both nutrients are very abundant in honey (35.8 g/100 g of glucose and 40.9 g/100 g of fructose), and have low abundance in spinach (0.11 g/100 g of glucose and 0.15 g/100 g of fructose). In contrast, protein and fiber have a strongly negative correlation in their amounts across foods ($r = -0.58$, $P = 5.6 \times 10^{-31}$). In the network, we also observed synergistic bottleneck nutrients that are linked to each other, such as choline and cholesterol (previously discussed, $r = 0.65$ and $P = 1.1 \times 10^{-25}$) or choline and linoleic acid (both favorable for the high NF in scallop, $r = -0.54$ and $P = 1.9 \times 10^{-6}$).

The existence of notably positive correlations in the network invites a closer examination of the connections between nutrients. Vitamins A and K have a highly positive correlation in their abundances across all foods ($r = 0.634$, $P = 3.2 \times 10^{-13}$). When correlations are measured within plant-derived and animal-derived foods separately, only plant-derived foods exhibit this positive correlation between vitamins A and K ($r = 0.632$ and $-0.13$ for plant- and animal-derived foods, respectively). Indeed, vitamins A and K are known to be synthesized in plants from a common molecular precursor, geranylgeranyl diphosphate [20]. Additionally, in our network, protein is one of the strongest hubs associated with many micronutrients, including choline and niacin. Protein and choline have a positive correlation not only across all foods ($r = 0.77$, $P = 4.0 \times 10^{-30}$) but also for plant-derived and animal-derived foods separately. The examination of each subgroup within animal-derived foods still reveals positive correlations between protein and choline (Supplementary Information). This connection between protein and choline remains valid, even when we remove the potential

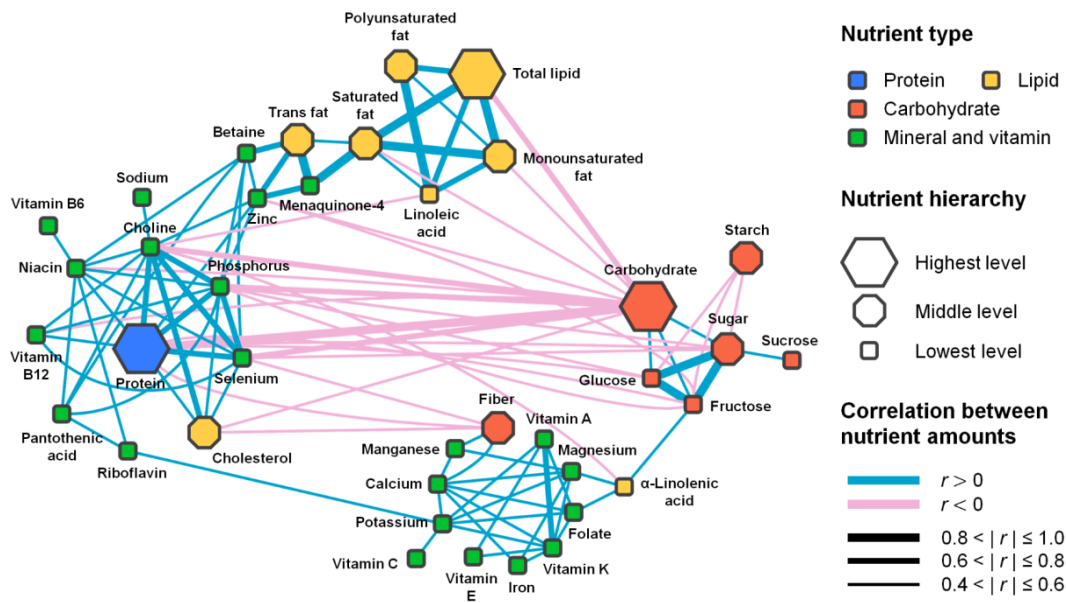

**Fig. 4. The nutrient-nutrient network.** Each node represents a nutrient, and the nodes are connected through correlations between the abundances of nutrients across all foods. The network is composed of three major groups of nutrients that are densely connected to one another through positive correlations. Between groups, nutrients have only sparsely positive or frequently negative correlations (Supplementary Information): the top and left side is for the first group, the right side is for the second group, and the bottom side is for the third group. Each node is colored according to the nutrient type. The shape of each node indicates the hierarchical or 'taxonomic' level of a nutrient, from 'Highest' (a general class of nutrients) to 'Lowest' (a specific nutrient). The color and thickness of each link correspond to the sign and magnitude of the correlation, respectively. Here, we only show the significant nutrients and correlations described in Supplementary Information, and we omit seven nutrients which don't have significant correlations with any others. We also omit amino acids because their correlations with other nutrients are very similar to the correlations of the total protein with others (thus, these correlations are redundant for visualization)

indirect causes of their correlation, such as the effects of phosphorus and cholesterol (compounds that have positive correlations with both protein and choline. See Supplementary Information). All of these results consistently support the robust association between protein and choline, although the detailed biological origins need to be elucidated. Similarly, protein and niacin have a highly positive correlation across all foods ($r = 0.59$, $P = 6.3 \times 10^{-26}$), and this correlation remains valid when measured within individual subgroups of foods separately (Supplementary Information). Tryptophan can be converted to niacin in animal livers [21], and this fact may contribute, at least in part, to such robust connection between protein and niacin. Interestingly, *trans*-fatty acids, which are famous for their associated risk of coronary heart disease [22], were found to have a highly positive correlation with zinc across all foods ($r = 0.62$, $P = 9.1 \times 10^{-9}$). Because *trans*-fatty acids also have a very positive correlation with saturated fatty acids ($r = 0.59$, $P = 1.4 \times 10^{-6}$), we considered the possibility that zinc may be indirectly correlated with *trans*-fatty acids through saturated fatty acids. By controlling for such an indirect effect, we found that, as long as the saturated fatty acid content of foods is at least 5.8 g/100 g (dry weight), zinc and *trans*-fatty acids still exhibit a highly positive correlation in their amounts without an indirect effect from the saturated fatty acids. Considering the effects from other than saturated fatty acids also did not impair the correlation between zinc and *trans*-fatty acids (Supplementary Information). This robust association between zinc and *trans*-fatty acids allows us to envision a potential biochemical mechanism that connects the two compounds. To the best of our knowledge, studies that mechanistically connect zinc and *trans*-fatty acids are not yet available, although other metal catalysts, such as copper and nickel, are known to facilitate the synthesis of *trans*-fatty acids [23].

The diversity of these pair-wise nutrient connections, previously discussed, raises the question of whether particular nutrients are bound coherently as underlying patterns for nutrient combinations in foods. Through the global examination of the nutrient-nutrient network, we identified three major groups of nutrients densely connected to each other through positive correlations, whereas between groups, the nutrients have only sparsely positive or frequently negative correlations (Fig. 4). The first group contains components of protein and lipid, which are seamlessly connected with a number of micronutrients such as phosphorus, selenium, zinc, choline, and niacin. The second group comprises digestible carbohydrates such as glucose and fructose. The third group consists of fiber, α-linolenic acid, and various micronutrients including vitamins A and K, folate, iron, and calcium. We observe that each of these three nutrient groups largely captures the nutrient characteristics of a specific food partition or category. Nutrients of animal-derived foods are highly enriched in the first group of nutrients, whereas those of plant-derived, low-calorie foods are enriched in the third nutrient group. The nutrient contents of the fat- and protein-rich foods within the plant-derived food partition are based on both the first and third nutrient groups. Furthermore, the nutrients of the carbohydrate-rich foods were found to primarily belong to the second and third nutrient groups. One may suppose that these results can be readily expected from the definitions of the food categories themselves, e.g., carbohydrate-rich foods, by definition, harbor large proportions of total carbohydrates. Our results, however, did not substantially change after controlling for such trivial or redundant factors related to macronutrients (Supplementary Information). This finding suggests that the network substructures themselves are the fundamental units of the underlying patterns for nutrient combinations in foods. Therefore, the global network of nutrients harbors a diverse repertoire of nutrient-to-nutrient connections that serve as building blocks for emerging characteristics, such as the characteristics that can distinguish different food partitions or categories.

## Conclusions

In this study, we have developed a unique computational framework for the systematic analysis of large-scale food and nutritional data. The networks of foods and nutrients offer a global and unbiased view of the organization of nutritional connections, as well as enable the discovery of unexpected knowledge regarding associations between foods and nutrients. Nutritional fitness, which gauges the quality of a raw food according to its nutritional balance, appears to be widely dispersed over different foods, raising questions on the origins of such variations between foods. Remarkably, this nutritional balance of food does not solely depend on the characteristics of individual nutrients but is also structured by intimate correlations among multiple nutrients in their amounts across foods. This underscores the importance of nutrient-nutrient connections, which constitute the network structures embodying multiple levels of the nutritional compositions of foods. Extending our analysis beyond raw foods to cooked foods is necessary to truly understand the nutritional landscape of the foods we consume daily (and is left for further study); however, considering only raw foods was sufficient to draw primary insights from a relatively simple system.

A number of applications would become achievable if the concepts presented here are judiciously combined with other practical approaches. The incorporation of region-specific information in our analysis may help design strategies for international food aid [24]. To develop such strategies, one can consider the prioritization of regional foods based on nutritional fitness, suggestions for locally-available dietary substitutes from a food-food network, the fortification of foods using bottleneck nutrients, and so forth. Our study also has implications for personalized nutrition [25]. People of different ages, genders, body compositions, health states, and physical activity levels can obtain their condition-specific information through our method, by simply adjusting the required calorie and nutrient intakes when generating irreducible food sets (Supplementary Information). The resulting irreducible food sets allow one to compute the nutritional fitness and bottleneck nutrients. This information can be of particular interest to individuals with certain dietary requirements, such as pregnant women, who are recommended to take more nutrients, e.g., essential amino acids and vitamins, than non-pregnant women [26]. On the other hand, it would be interesting to check how different farming methods for each food affect the food's nutritional composition and thereby its nutritional fitness. Currently, our data source does not provide such information about farming methods (Supplementary Information). Furthermore, considerations of food taste and financial, seasonal, and cultural factors in our analysis may improve the applicability of our methods towards nutritional policy making, nutrition education, and food marketing [1, 27, 28], as well as the aforementioned food aid and personalized nutrition. Finally, our systematic approach sets the foundation for future endeavors to enhance the understanding of food and nutrition.

## Acknowledgments

We thank Flora Aik for assistance with the data preparation and William Helferich, Manabu Nakamura, Juan Andrade, Yong-Yeol Ahn, Sanguk Kim, and Soo-Yeun Lee for useful discussions. This work was supported by the Basic Science Research Program through the National Research Foundation of Korea (NRF-2012R1A1A2008925).